\documentclass{sigchi}

\CopyrightYear{2019}
\setcopyright{rightsretained}
\conferenceinfo{UIST'19,}{October  20--23, 2019, New Orleans, LA, USA}

\usepackage{balance}       
\usepackage{graphics}      
\usepackage[T1]{fontenc}   
\usepackage{txfonts}
\usepackage{mathptmx}
\usepackage[pdflang={en-US},pdftex]{hyperref}
\usepackage{color}
\usepackage{booktabs}
\usepackage{textcomp}

\usepackage{microtype}        
\usepackage{ccicons}          

\usepackage{todonotes}

\def\plaintitle{Homo Cyberneticus: The Era of Human-AI Integration}

\def\emptyauthor{}
\def\plainkeywords{HCI vision; human-augmentation; human-AI integration}

\makeatletter
\def\url@leostyle{%
  \@ifundefined{selectfont}{
    \def\UrlFont{\sf}
  }{
    \def\UrlFont{\small\bf\ttfamily}
  }}
\makeatother
\def\pprw{8.5in}
\def\pprh{11in}

\definecolor{linkColor}{RGB}{6,125,233}
\hypersetup{%
  pdftitle={\plaintitle},
  pdfauthor={\emptyauthor},
  pdfkeywords={\plainkeywords},
  pdfdisplaydoctitle=true, 
  bookmarksnumbered,
  pdfstartview={FitH},
  colorlinks,
  citecolor=black,
  filecolor=black,
  linkcolor=black,
  urlcolor=linkColor,
  breaklinks=true,
  hypertexnames=false
}

\newcommand{\commentout}[1]{}

\begin{document}

\title{\plaintitle}

\numberofauthors{1}
\author{%
  \alignauthor{Jun Rekimoto\\
  \email{The University of Tokyo, Sony CSL, http://lab.rekimoto.org, rekimoto@acm.org}}\\
}

\maketitle

\commentout{
\begin{abstract}
abstract
\end{abstract}
}


\commentout{

}


\commentout{
Please use the 2012 Classifiers and see this link to embed them in the text: \url{https://dl.acm.org/ccs/ccs_flat.cfm}
}

\section{Human-Augmentation}

\vspace{0.1
cm}
\begin{quotation}
\noindent
{\it
Neo: Can you fly that thing?\\
Trinity: Not yet.\\
}
\end{quotation}

\vspace{-0.4cm}

In the movie ``The Matrix,'' Trinity responds to Neo right before having the helicopter's maneuverability downloaded into her brain~\cite{matrix}. Will such a future come? Although it may take some time to reach the world of The Matrix, technologies that can enhance and augment human abilities are advancing rapidly~\cite{Rekimoto1995uist,Rekimoto:2014:NYA:2669485.2670531,7842855}.

The idea that technology enhances humanity has a long history.  Robert Hooke, the inventor of the optical microscope, considered the microscope to be an extension of vision, and that {\it ``There may be found many Mechanical Inventions to improve our other Senses, of hearing, smelling, tasting, touching.''}~\cite{micrographia}. Vannevar Bush's MEMEX was intended to EXpand human MEMory~\cite{bush45}.  
For Marshall McLuhan, all media technologies were extensions of humans, as radio was an ear extension and television was an eye extension~\cite{macluhan}.
Douglas Engelbart explained {\it ``The mouse was just a tiny piece of a much larger project aimed at augmenting human intellect.''}~\cite{engelhypo,engelbart62}.  His ultimate goal was to augment human intellect, and mice and GUIs were tools to realize that goal. However, the results of Engelbart and subsequent Alto~\cite{Kay:1977:PDM:1300728.1301186} were so successful that our research interests have been more directed towards improving such tools, interaction technologies and input/output devices, than exploring and deepening the various possibilities of enhancing humans.

In that regard, J.C.R. Licklider's ``Man-Computer Symbiosis''~\cite{licklider60} is worth reviewing. 
Here, {\it symbiosis} means 
{\it ``living together in intimate association, or even close union, of two dissimilar organisms.''}
The assumption is that humans and computers compliment each other with their respective abilities.
More interestingly, Licklider also gives two examples of {\it non-symbiotic} human-computer relationships. First, the concept of the {\it mechanically extended man} is not symbiotic. The concept refers to something like a prosthetic leg or glasses, but it can now be said that it leads to the development of cyborgs. The other non-symbiotic example is {\it humanly extended machines}, assuming humans are participating as operators in automated factories. However, this can now be seen as when human abilities are given to machines, such as a telepresence robot~\cite{waldo,minsky80omni}. Systems that learn from gathering collective human intelligence, such as recommendation systems or machine translation systems, may fall into this category. Thus, Licklider's counterexamples of symbiosis can be said to show two important directions of human augmentation.


\section{Human-AI Integration}

One of my current projects is a silent voice system that generates a voice from intraoral movements taken from ultrasound echo images using deep-neural networks~\cite{Kimura:2019:SUI:3290605.3300376}.
The project showed the possibility of verbal communication in situations where vocal speech is difficult, or when someone has vocal cord damage. 

This human-augmenting system is also a kind of training system.
The system learns to generate voices and helps users improve their silent lip-mouse-moving patterns by listening to the sounds that the system generates. 
Glove Talk II~\cite{Fels:1995:GAG:223904.223966} is a voice generation system that uses hand gestures. The user's hand gestures and voice synthesizer parameters are connected with neural networks. 
A user can learn to control voices with hand gestures and can even utter non-verbal expressions such as ``Yeah?'' with their hands. 
These examples are similar to how people learn skills.
When we learn utterances, the coordination of the motor cortex that drives the oral cavity, tongue, vocal folds,
and the auditory cortex forms a tight feedback loop to obtain a better speech performance.
The examples above extend this loop and connects 
organic neural networks (e.g., the brain) and artificial neural networks in the one loop.

I think there are many possibilities in this direction.  
Rather than considering AI an autonomous or separated entity, 
we may be able to regard AI as an integrated part of humans. 
Human motor ability and cognitive ability can be expanded by AI closely linked with humans.
I would like to name this {\it human-AI integration} rather
than human-AI interaction (see the Appendix diagram for the four categories of HCI and AI relationships). 
As an extension of such research, the combination of an invasive brain and AI will also be considered~\cite{Musk703801}. What happens if humans form a network through these connections~\cite{brainnet}? If such a connection can directly extend our capabilities, we will be able to re-design ourselves, and it will be the future of {\it interaction design}.

Arthur Charles Clarke stated, ``The old idea that man invented tools is a misleading half-truth. It would be more accurate to say that {\it tools invented man}''~\cite{clarke}. A major theme of future HCI research will be reinventing humans in an era when humans and technology integrate.



\balance

\begin{figure*}[t]
    \centering
    \includegraphics[width=0.7\textwidth]{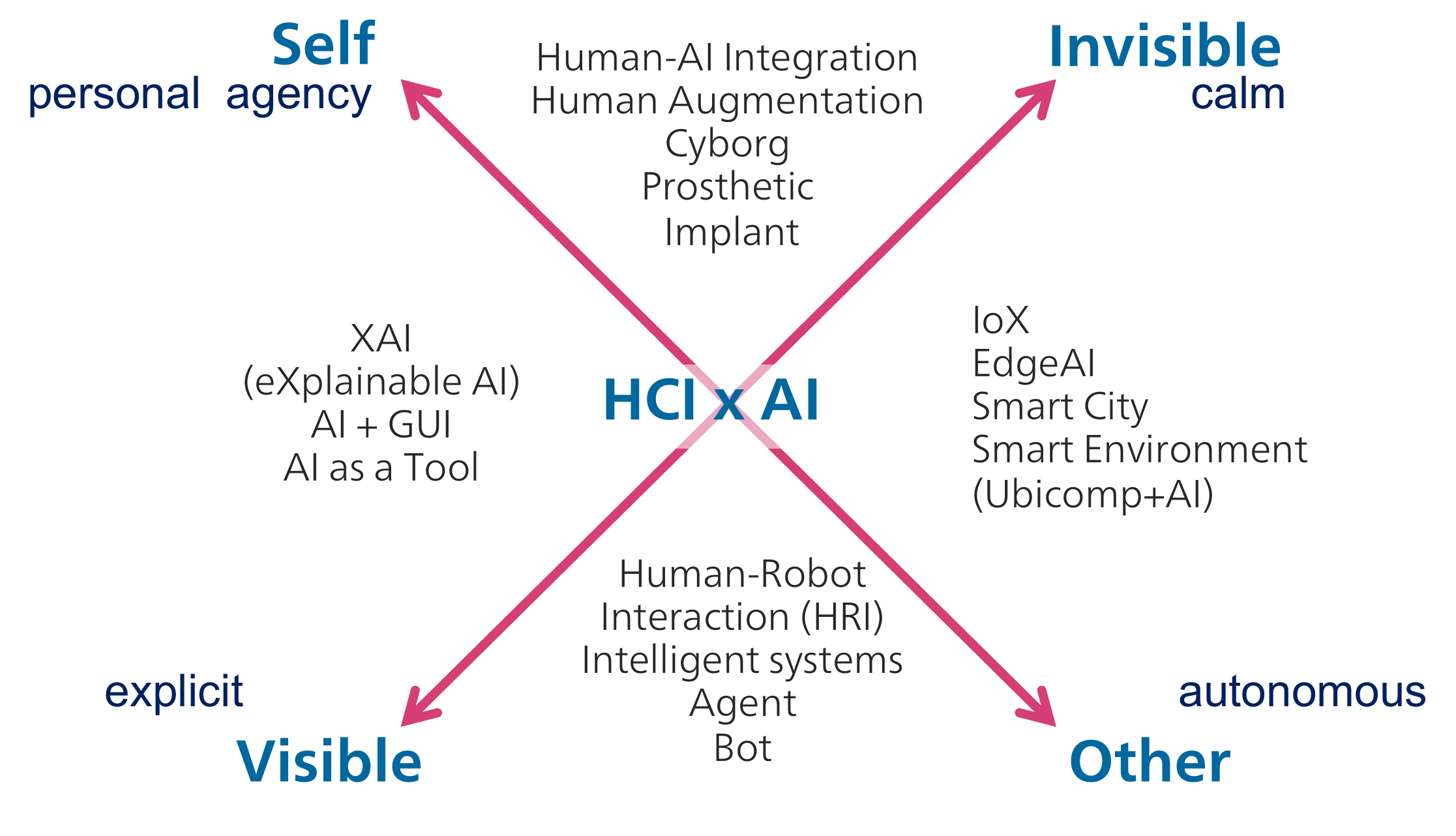}
    \caption{Appendix. Relationships of HCI and AI: The relationship between AI and humans is classified by two axes. The one axis is visible or invisible; it shows whether an interaction is explicit or a more integrated and invisible (calm) interaction.
    The other axis is self or other; it classifies by interactions with individuals or by interactions with the outside world. As shown in this diagram, HCI and AI can have four quadrants.}
    \label{fig:my_label}
\end{figure*}

\bibliographystyle{SIGCHI-Reference-Format}
\bibliography{book}

\balance

\end{document}